\documentclass[12pt, doublespaced]{article}

\usepackage{graphicx}
\usepackage{epstopdf}
\usepackage{epsfig}
\usepackage{subfigure}
\usepackage{amsmath}
\usepackage{amssymb}

\def\lesssim{\mathrel{\rlap{\lower4pt\hbox{\hskip1pt$\sim$}}}<}
\def\gtrsim{\mathrel{\rlap{\lower4pt\hbox{\hskip1pt$\sim$}}}>}
\newcommand{\sv}{\langle\sigma v\rangle}

\newcommand{\msun}{M_{\odot}}

\def\be{\begin{equation}}
\def\ee{\end{equation}}

\begin{document}

\title {Supermassive Dark Stars: Detectable in JWST}
\author{Katherine Freese$^1$, Cosmin Ilie$^1$, 
Douglas Spolyar$^2$, Monica Valluri$^3$,\\ and Peter Bodenheimer$^4$}

\date{}
\maketitle

\begin{center}
$^1$Michigan Center for Theoretical Physics, Department of Physics,
University of Michigan, Ann Arbor, MI 48109\\ 
$^2$Center for Particle Astrophysics, 
Fermi National Accelerator Laboratory, Batavia, IL 60510\\
$^3$ Department of Astronomy, University of Michigan, Ann Arbor, MI 48109
$^4$ UCO/Lick Observatory, Dept. of Astronomy and Astrophysics, University of California, 
Santa Cruz, CA 95064
\end{center}

\begin{abstract}

The first phase of stellar evolution in the history of the Universe 
may be Dark Stars, powered by dark matter heating rather than by nuclear 
fusion. Weakly Interacting Massive Particles, which may be their own
 antipartners, collect inside the first stars and annihilate to produce a heat 
source that can power the stars for millions to billions of years. 
In this paper we show that these objects
can grow to be supermassive dark stars (SMDS) with masses
 $\gtrsim (10^5-10^7) \msun$.  The growth continues as long as dark
matter heating persists, since dark stars are large and cool (surface
temperature $\lesssim 5\times 10^4$K) and do not emit enough ionizing
photons to prevent further accretion of baryons onto the star.  The
dark matter may be provided by two mechanisms: (1) gravitational
attraction of dark matter particles on a variety of orbits not previously considered, and (2)
capture of WIMPs due to elastic scattering.
Once the dark matter fuel is exhausted, the SMDS becomes a heavy 
main sequence star; these stars eventually collapse to form massive black
holes that may provide seeds for supermassive black holes in the Universe.
SMDS are very bright, with luminosities exceeding
$(10^9-10^{11}) L_\odot$. We demonstrate that for several reasonable
parameters, these objects will be detectable with JWST.  Such an
observational discovery would confirm the existence of a new phase of
stellar evolution powered by dark matter.

\end{abstract}

\section{Introduction}

Spolyar et al. (2008) first considered the effect of dark matter (DM)
particles powering the first stars.  The first stars formed when the
universe was about 200 million years old, at $z=10-50$, in $10^6
M_\odot$ haloes consisting of $85\%$ DM and $15\%$ baryons in the form
of H and He from big bang nucleosynthesis; for reviews of the standard
picture of the formation of the first stars, see Barkana \& Loeb
(2001), Yoshida et al. (2003), Bromm \& Larson (2004), and Ripamonti
\& Abel (2005).  The canonical example of particle DM is Weakly
Interacting Massive Particles (WIMPs). In many theories WIMPs are
their own antiparticles and annihilate among themselves wherever the
DM density is high.  Recently there has been much excitement in the
dark matter community about possible detections of WIMPs via
annihilation to positrons seen by the PAMELA satellite (Adriani et
al. 2009); annihilation to $\gamma$-rays seen by FERMI (Abdo et
al. 2009, Dobler et al. 2009), and in the direct detection experiments
DAMA and CDMS (Bernabei et al. 2009, Ahmed et al. 2009).  The
annihilation rate is $n_\chi^2 \langle \sigma v \rangle$ where
$n_\chi$ is WIMP density and we take the standard annihilation cross
section\footnote{Annihilation in the early universe with this value of
the cross section leaves behind the correct relic WIMP DM abundance
today, $\sim 24$\% of the energy density of the universe.}
\begin{equation}
\label{eq:sigmav}
\langle \sigma v \rangle = 
3 \times 10^{-26} {\rm cm}^3/{\rm s},
\end{equation}
and WIMP masses in the range 1 GeV-10 TeV.  The first stars are
particularly good sites for annihilation: they form in the right place
--- in the high density centers of DM haloes --- and at the right time
--- at high redshifts (density scales as $(1+z)^3$).
Spolyar et al. (2008) found that above a certain density ($\approx
10^{13}$ cm$^{-3}$) the WIMP annihilation products remain trapped in
the star, thermalize with the star, and thereby provide a heat source.
These first Dark Stars (DS) are stars made primarily of hydrogen and
helium with only $\sim 0.1$\% of the mass in the form of DM; yet they
shine due to DM heating. Note that the term 'Dark' refers to the power
source, not the appearance or the primary matter constituent of the
star.

Dark stars are born with masses $\sim 1 \msun$.  They are giant puffy
($\sim 10$ AU), cool (surface temperatures $<10,000$K), yet bright
$>10^6 L_\odot$ objects (Freese et al. 2008a).  They reside in a large
reservoir ($\sim 10^5 \msun$) of baryons, i.e., $\sim 15$\% of the
total halo mass.  These baryons can start to accrete onto the dark
stars.  Previous work (Freese et al. 2008a; Spolyar et al. 2009)
followed the evolution of dark stars from their inception at $1
\msun$, as they accreted baryons from the surrounding halo, up to
$\sim 1000 \msun$.  Dark stars can continue to grow in mass as long as
there is a supply of DM fuel.
 
 The exciting new development of this paper is that we follow the
 growth of dark stars to become supermassive dark stars (SMDS) of mass
 $M_* >10^5 \msun$.  
  Specifically we study the formation of $10^5 \msun$ SMDS  in $10^6 \msun$
DM haloes and $10^7 \msun$ SMDS  in $10^8 \msun$ haloes;
perhaps the SMDSs become even larger.
 Hoyle \& Fowler (1963) first postulated the
 existence of such large stars but were not aware of a mechanism for
 making them.  Now the confluence of particle physics with
 astrophysics may be providing the answer.  The key ingredient that
 allows dark stars to grow so much larger than ordinary fusion powered
 Population III stars is the fact that dark stars are so much cooler.
 Ordinary Pop III stars have much larger surface temperatures in
 excess of 50,000K. They produce ionizing photons that provide a
 variety of feedback mechanisms that cut off further accretion.  McKee
 \& Tan (2008) have estimated that the resultant Pop III stellar
 masses are $\sim 140\msun$.  Dark stars are very different from
 fusion-powered stars, and their cooler surface temperatures allow
 continued accretion of baryons all the way up to enormous stellar
 masses, $M_* > 10^5\msun$. 

WIMP annihilation produces energy at a rate per unit volume 
\begin{equation}
\hat Q_{DM} = n_\chi^2 \langle \sigma v \rangle m_\chi =
\langle \sigma v \rangle \rho_\chi^2/m_\chi ,
\label{eq:Q}
\end{equation}
where $n_\chi$ is the WIMP number density, $m_\chi$ is the WIMP mass,
and $\rho_\chi$ is the WIMP energy density.  The annihilation products
typically are electrons, photons, and neutrinos. The neutrinos escape
the star, while the other annihilation products are trapped in the
dark star, thermalize with the star, and heat it up.  The luminosity
from the DM heating is
\begin{equation}
\label{DMheating}
L_{DM} \sim f_Q \int \hat Q_{DM} dV 
\end{equation}
where $f_Q$ is the fraction of the annihilation energy deposited in
the star (not lost to neutrinos) and $dV$ is the volume element. We
take $f_Q=2/3$ as is typical for WIMPs.

Typically $ (100-10^4) \msun$ of dark matter (up to $\sim 1\%$ of the halo mass) 
must be consumed by the
star in order for large SMDS masses $M_* \sim 10^5\msun$ to be
reached. We will consider two different scenarios for supplying this
amount of dark matter: \hfil
\break
\vskip 0.1 in
\noindent
  1)  Extended Adiabatic Contraction, labeled ``without
 capture'' below. In this case, dark matter is supplied by the
 gravitational attraction of the baryons in the star. The amount of DM
 available for DM annihilation due to adiabatic contraction may be
 larger than our previous estimates which were based on the assumption
 that DM halos are spherical.  In a non-spherical DM halo, the supply of DM 
available to the star can be considerably
 enhanced, as we discuss in more detail in \S \ref{DMdens}.
In any case, this mechanism
relies solely on the particle physics of WIMP annihilation and does not include capture of DM by baryons (discussed below).
 
  \hfil\break 

2) Extended Capture, labeled ``with capture'' below.
  Here, the star is initially powered by the DM from adiabatic
  contraction (AC), but the AC phase is taken to be short $\sim
  300,000$ years; once this DM runs out the star shrinks, its density
  increases, and subsequently the DM is replenished inside the star by
  capture of DM from the surroundings (Freese et al. 2008b; Iocci
  2008) as it scatters elastically off of nuclei in the star.  In this
  case, the additional particle physics ingredient of WIMP scattering
  is required.  This elastic scattering is the same mechanism that
  direct detection experiments (e.g.  CDMS, XENON, LUX, DAMA) are
  using in their hunt for WIMPs.  In previous work (Freese et
  al. 2008a; Spolyar et al. 2009), we assumed minimal capture, where DM
  heating and fusion contributed equally to the luminosity once the
  star reached the main sequence. Here, we consider the more sensible
  case where DM heating dominates completely due to larger ambient DM
  density, and the star can grow to become supermassive.

 Supermassive dark stars can result from either of these mechanisms
 for DM refueling inside the star.  The SMDS can live for a very long
 time, tens to hundreds of million years, or possibly longer (even to
 today). We find that $\sim 10^5 \msun$ SDMSs are very bright $\sim 3
 \times 10^9 L_\odot$ which makes them potentially observable by James
 Webb Space Telescope (JWST).  We also note that SMDS may become even
 more massive a) if they form in larger haloes or b) the DM haloes in
 which they initially form merge with other haloes so that there is
 even more matter to accrete (n.b. alternatively these mergers may
 remove the DS from their high DM homes and stop the DM heating).  For
 example, if dark stars form in $10^8 \msun$ haloes, then they could
 in principle grow to contain all the baryons in the halo, i.e. $M_* >
 10^7 \msun$.  Since the luminosity scales as $L_* \propto M_*$ these
 SMDS would be even brighter, $L_* > 10^{11} L_\odot$ and are hence
 even better candidates for discovery in JWST.
 
 Once the SMDS run out of DM fuel, they contract and heat up. The core
 reaches $10^8 $K and fusion begins.  As fusion-powered stars they
 don't last very long before collapsing to black holes.  Again, this
 prediction is different from standard Pop III stars, many of which
 explode as pair-instabilty supernovae (Heger \& Woosley 2002) with
 predicted even/odd element abundance ratios that are not (yet)
 observed in nature.  The massive black holes (BH) remnants of the
 SMDSs are good candidates for explaining the existence of
 $10^9~\msun$ BHs which are the central engines of the most distant ($z>5.6$)
 quasars in the Sloan Digital Sky Survey (SDSS) (Fan et al. 2001,
 2004, 2006) 
\footnote{We thank N. Yoshida for pointing this out to us}.
 
 The idea of supermassive DS and the resultant $> 10^5 \msun$ BH was
 originally proposed by Spolyar et al. (2009).  Subsequently Umeda et
 al. (2009) took their existing stellar codes and added DM
 annihilation to allow the mass to grow. They started from Population
 III stars in which fusion was already present, assuming they then
 encounter a reservoir of DM.  Our work differs in that we start from
 the very beginning with collapsing protostellar clouds that
 transition into dark stars, which can be DM powered for millions to
 billions of years before fusion ever begins.  Our SMDSs are truly
 primordial supermassive stars.

Begelman (2009) presents another alternative for the formation of
supermassive stars: rapid accretion onto stars which already have
hydrogen burning in them. His ``quasistars'', another possible route
to large BH, are quite different from the SDMS discussed in this
paper.

Various other authors that have explored the repercussions of DM
heating in the first stars (Taoso et al. 2008; Yoon et al. 2008;
Ripamonti et al. 2009; Iocco et al. 2008). The possibility that DM
annihilation might have effects on {\it{ today's}} stars was initially
considered in the $'80$s and early $'90$s (Krauss et al. 1985; Bouquet
\& Salati 1989; Salati \& Silk 1989; Graff \& Freese 1996) and has
recently been studied in interesting papers by Moskalenko \& Wai
(2007), Scott et al. (2007), Bertone \& Fairbairn (2007) and Scott et
al. (2009).  Other constraints on DS will arise from cosmological
considerations.  A first study of their effects (and those of the
resultant MS stars) on reionization has been done by Schleicher et
al.(2008, 2009) and further work in this direction is warranted.

In this paper we examine the SMDS that result from the two mechanisms
discussed above for DM refueling inside the star.  In Section II we
discuss the procedure for calculation of models; in Section III we
present results; and in Section IV we end with a discussion.

\section{Structure and Evolution of the Dark Star}

DM heating is very different from fusion. In order to overcome the
Coulomb barriers between nuclei, fusion requires very high
temperatures and densities in the star.  Fusion is not very efficient
in that only $<1\%$ of the nuclear mass is converted to heat. WIMP
annihilation, on the other hand, takes place at high dark matter
densities regardless of the temperature.  It is almost 100\% efficient
since O(1) of the WIMP mass is converted to useful energy.  Thus, in
the evolution of the first protostars, DM heating becomes important
early. Here we start the calculation when the DS is massive enough (3
M$_\odot$) so that it is in hydrostatic equilibrium and most of the
hydrogen and helium is ionized.  The contribution to DM luminosity is
roughly constant as a function of radius throughout the DS, unlike
fusion which takes place only at the (high temperature) core of the
star.

\subsection{Basic Equations}\label{basiceq}

We use the numerical code previously discussed in detail by Freese et
al. (2008a) and Spolyar et al. (2009).  We make the assumption that a
DS can be described as a polytrope
\begin{equation}
P=K\rho^{1+1/n}
\end{equation}
in hydrostatic equilibrium. Here $P$ is the pressure, $\rho$ is the
density, and $K$ is a constant.  We solve the equations of stellar
structure with polytropic index $n$ initially 1.5, as appropriate for
convective stars, and made a gradual transition to $n=3$ as the star
becomes radiative in the later phases.  We require that at each
time-step during the accretion process the star is in hydrostatic
equilibrium, 
\be
\label{hydro}
 \frac{dP}{dr}=-\rho(r)\frac{GM_r}{r^2}
\ee 
where ${dM_r \over dr} = 4 \pi r^2 \rho(r)$,
and $M_r$ is the mass enclosed in a spherical shell of radius $r$. 
The equation of state includes radiation pressure, 
\be\label{eqstate}
P(r)=\frac{k_B\rho(r)T(r)}{m_u\mu}+\frac{1}{3}aT(r)^4\equiv P_g+P_{rad}
\ee
where $k_B$ is the Boltzmann constant, $m_u$ is the atomic mass unit and $\mu
=0.588$ is the mean atomic weight. The opacity is obtained from
a zero metallicity table from OPAL (Iglesias \& Rogers 1996)  supplemented at low
 temperatures by opacities from Lenzuni et al. (1991)  for $T<6000$K.   The further 
assumption is made that the radiated  luminosity of the star $L_\ast$ is balanced
by the rate of energy output by all internal sources, $L_{\rm tot}$, as described
 below in \S \ref{ensources}.
\be
L_*=4\pi R_*^2\sigma_BT_{eff}^4 = L_{\rm tot}.
\label{eq:thermal}
\ee
 where $T_{eff}$ is the surface temperature, and  $R_\ast$ is the
total radius.

Starting with a mass $M$ and an estimate for the outer radius $R_*$, the code integrates
 the structure equations                outward from the center.
 The total rate of energy production $L_{\rm tot}$ is compared to the stellar radiated
 luminosity, as in equation (\ref{eq:thermal}) and the radius is adjusted
 until the condition of  thermal equilibrium is met (a convergence of $1$ in $10^4$ is reached).  

\subsection{Dark Matter Densities}\label{DMdens}

We now describe the two different mechanisms for supplying the DM density.

{\it Extended Adiabatic Contraction:} 
As the baryons start to collapse
into a protostellar cloud at the center of the DM halo, the DM
responds to the changing gravitational potential well and falls in as
well.  As described in our previous work (Spolyar et al. 2008), we
will use adiabatic contraction (AC) to describe this increase in DM
density. For the case of spherical haloes, we previously found, by
performing exact calculations for comparison (Freese et al. 2009),
that the simple Blumenthal method (Blumenthal et al. 1985; Barnes \&
White 1984; Ryden \& Gunn 1987) gives reliable results for the final
DM densities up to an unimportant factor of two; others confirmed this
conclusion (Natarajan et al. 2009; Iocco et al. 2008; Sivertsson \&
Gondolo 2010).  Using this simple approach during the AC phase, we
found that $\rho_\chi \sim 5 {\rm GeV/cm}^3 (n_h/{\rm cm}^3)^{0.81}$
where $n_h$ is the gas density. These are the values we will use
during AC.

In our previous work, we probably underestimated the lifetime of the
DM inside the star due to AC.  In our previous work we treated the DM
halo as spherical and ran up the DS mass to the point where the DM
initially inside the star was entirely consumed by annihilation.  The
DS mass at this point is $O(10^3)\msun$ after a lifetime of $\sim
300,000$ years, and the amount of DM consumed has only amounted to
$\sim 1 \msun$.  In a spherical DM halo, the orbits of DM particles
are planar rosettes (Binney \& Tremaine 2008) conserving energy as
well as all three components of angular momentum; consequently the
central hole (or ``empty loss cone'') that results from DM
annihilation cannot be repopulated once it is depleted. (Note that
although DM annihilation creates a central hole in the DM density, the entire region
is filled with baryons and hence the potential is stable.)  However,
it is well known that DM halos formed in hierarchical structure
formation simulations are not spherical but are prolate-triaxial
(Bardeen et al. 1986; Barnes \& Efsthatiou 1987; Frenk et al. 1988;
Dubinski \& Carlberg 1991; Jing \& Suto 2002; Bailin \& Steinmetz
2005; Allgood et al. 2006) with typical axis ratios of
(short-axis)/(long-axis) $\sim 0.6-0.8$. In triaxial potentials, the
orbits do not conserve angular momentum. In particular there are two
families of ``centrophilic orbits'' (box orbits and chaotic orbits)
which oscillate back and forth through the potential and can travel
arbitrarily close to the center (Schwarzschild 1979, Goodman \&
Schwarzschild 1981; Gerhard \& Binney 1985; de Zeeuw 1985;
Schwarzschild 1993; Merritt \& Fridman 1986; Merritt \& Valluri 1996).
Unlike an orbit in a spherical potential which has a constant
pericenter radius (the distance of closest approach to the center of
the potential), the pericenter radius of a centrophilic orbit varies
over time extending from $r_{\rm peri} = [0, r_{\rm max}]$ where
$r_{\rm max}$ is sometimes referred to as the ``throat'' of the orbit
(Gerhard \& Binney 1985).  As any one particle traverses the center of
the dark star, it may indeed be removed from the pool by annihilation.
However, it was unlikely to pass through the star (of radius $r_*$) on
its next orbit anyhow since in general $r_{max} >> r_*$.  Instead, a
particle on a different ``centrophilic orbit'' enters the star for
the first time maintaining the steady state central DM density
cusp. Unlike in the case of a spherical DM halo, where annihilation
steadily depletes the central density cusp, in a triaxial halo there
is a high probability for a particle on a centrophilic orbit to pass
through the center for the first time on any given orbital
crossing. This is particularly true in potentials with central point
masses which are dominated by chaotic orbits and are therefore ergodic
(Merritt \& Valluri 1996).  Hence, the central DM density can remain
much higher than we previously expected.
 
 The dynamics of the refilling of the central ``loss cone'' in the
 case of spherical and non-spherical collisionless systems has been
 studied previously in the context of capture of stars by a central
 black hole (Gerhard \& Binney 1985; Magorrian \& Tremaine 1999;
 Merritt \& Poon 2004). The details of filling rate for the specific
 case of the DS will be presented separately in a later paper; however
 these previous papers showed that in non spherical systems the loss
 cone could remain full for a factor of 2 to $10^4$ times longer than
 in the spherical case depending on whether the potential was
 axisymmetric (Magorrian \& Tremaine 1999) or triaxial (Merritt \&
 Poon 2004), respectively.  Since DM halos are known to be triaxial
 this suggests that the duration for which the central hole remains
 full (i.e. has orbits with the low angular momentum necessary for
 annihilation) can increase from 300,000 years to as much as
 $3\times 10^9$ yrs possibly
 allowing the DS to be detectable by JWST.

 These more optimistic estimates require that a significant fraction
 of the orbits in these early DM halos are chaotic and boxlike. One
 important potential concern with assuming conditions in a triaxial
 halo is that several studies have shown that the growth of central
 baryonic components tend to make DM halos more axisymmetric than in
 purely dissipationless simulations (Dubinski 1994, Evrard et
 al. 1994, Merritt \& Quinlan 1998, Kazantzidis et al. 2004,
 Debattista et al. 2008, Tissera et al. 2009), and axisymmetric models
 are generally not expected to contain centrophilic orbit
 families. However, Valluri et al. (2010) recently showed that when a
 compact central baryonic component is grown adiabatically inside a
 triaxial DM halo,  the final halo that results from
 the adiabatic growth of such a baryonic component looks nearly
 oblate, yet its orbit population can contain a significant fraction of
 centrophilic orbits, since box orbits preferentially deform their
 shapes rather than converting to centrophobic tube families.  Furthermore they showed
 that a significant fraction of the orbits (both box orbits and tube orbits which were traditionally thought to avoid the center) experience
 strong chaotic scattering, a mechanism that could drive them close to
 the center. They  found that for a fixed ratio of the mass of the baryonic component to
 mass of the DM halo ($M_*/M_{\bf DM} = 10^{-3}$), the smaller the
 radius of the baryonic component ($r_*$), relative to the virial
 radius of the DM halo ($r_{\rm vir}$) the greater was the fraction of
 chaotic orbits. The most compact baryonic component studied by them
 had $r_*/r_{\rm vir} = 4.6 \times 10^{-4}$, which is significantly
 larger than that for our fiducial SMDS for which $r_*/r_{\rm vir} = 2
 \times 10^{-7}$. The compactness of the baryonic component in the DS
 relative to its halo is important since Valluri et al. (2010) showed
 that when the central component became very compact, orbits that were
 previously thought to be immune to becoming chaotic (the ``long-axis
 tubes'' which are important in prolate DM halos) also become chaotic.
 Valluri \& Freese (2010) are currently computing the rate at which
 the ``loss cone'' at the center of the SDMS will be refilled for a
 range of possible halo and DS masses.
 
 It is interesting to speculate that the Initial Mass Function of the
first stars may be determined by the cutoff of the DM supply, which
will take place at different SMDS masses in different haloes,
depending on the details of the cosmological merger history.  As the
SMDS mass becomes a significant fraction of the halo mass ($\sim
5-10$\%), it can significantly affect the shape of the halo, causing
it to become more axisymmetric (Merritt \& Quinlan 1998) and thereby
potentially cutting off the DM supply; on the other hand, in the
meantime the halo is growing larger due to mergers, which will
replenish the population of radial orbits.
 Numerical simulations with better resolution
 than currently possible will be necessary to address these questions.

In the meantime, in our case without capture, we assume that the
required DM is present and allow the stellar mass to grow to the point
where most of the baryonic content of the initial halo is inside a
single SMDS. 

The amount of dark matter required inside the star to
sustain long enough DM heating to reach a $10^5 \msun$ SMDS
is still small, $\sim 100\msun$ for
accretion rate $\dot M = 10^{-2}\msun$/yr and $\sim 10^4 \msun$ for
accretion rate $\dot M = 10^{-3}\msun$/yr, out of a total
$10^6\msun$ halo.  In the code we accomplish this by not removing
annihilated DM from the pool.  More precise studies must be performed
later in which we follow individual particle orbits in triaxial potentials to better
determine the precise DM density at any one time.

{\it Extended Capture:} In our model labeled ``with capture'', we
assume (as in our previous papers) that the DM due to adiabatic
contraction runs out in $\sim 300,000$ yrs.  For a while DM heating
becomes unimportant and the DS has to contract to maintain pressure
support.  Then DS is refueled in the later stages due to capture of
further DM from the ambient medium. This refueling requires an
additional piece of particle physics: scattering of DM off the nuclei
inside the star.

Some of the WIMPs from the ambient medium that have orbits passing
through the star will eventually be captured and sink to the center,
where they can annihilate efficiently.  The capture process is
irrelevant during the early evolutionary stage of the DS, since the
baryon density is not high enough at that point, leading to very small
scattering probabilities.  However, once the DS approaches the main
sequence, the baryon densities become high enough for substantial
capture to be possible.  This mechanism was first noticed by Freese et
al. (2008b) and Iocco (2008).

In our previous work (Spolyar et al. 2009) we investigated a 'minimal
capture' case which did not cause the DS to grow much more massive
than the original case without capture $\sim 1000 \msun$; but we
stated our intention to work out the case of a more substantial
background DM density in which case the DS would end up supermassive.
This is what we investigate here.  The capture rate is sensitive to
the product of two uncertain quantities: the scattering cross section
of WIMP interactions with the nuclei $\sigma_c$ and the ambient DM
density $\bar\rho_\chi$.  Since the capture mechanism depends only on
the product of these two quantities, one can interchangeably vary
either of these. For illustration purposes we will fix
$\sigma_c=10^{-39}~{\rm cm}^2$ and vary $\bar\rho_\chi = (10^{10} -
10^{14}) {\rm GeV/cm}^3$.  The latter quantity is the largest
reasonable amount based on our results for AC at the DS surface; and
the former is the ``minimal capture'' value considered in all our
previous papers.
Our fiducial cross section is just below the experimental bound for
spin-dependent (SD) scattering; the bound on spin-independent (SI)
scattering is much tighter: $\sigma_{c,SI} <3.8 \times
10^{-44}$~cm$^2$ for $m_\chi = 100$GeV (Ahmed et al. 2009). We will
show that capture can produce sufficient DM in the star to keep DM
heating alive for a long time. The details of our procedure for including capture have
previously been presented in Spolyar et al. (2009) and will not be
repeated here.

\subsection{Energy Sources:}\label{ensources}
There are four possible contributions to the DS luminosity:
\be
L_{tot}=L_{DM}+L_{grav}+L_{nuc}+L_{cap}
\ee
from DM annihilation, gravitational contraction, nuclear fusion, and captured DM respectively.
The heating due to DM annihilation in Eqns. (\ref{eq:Q}) and
(\ref{DMheating}) dominates from the time of DS formation until the
adiabatically contracted DM runs out.  As described previously, in our
``without capture'' models this stage never ends due to extended
adiabatic contraction.  In the models ``with capture'', on the other
hand, we take this phase to end after $\sim$ 300,000 years, so that
the DS has to contract in order to maintain pressure support.  The
contribution $L_{grav}$ due to gravitational energy release is
calculated as in Spolyar et al. (2009).
As the DS contracts, the density and temperature increase to the point
where nuclear fusion begins.  We include deuterium burning starting at
$T \sim 10^6K$, hydrogen burning via the equilibrium proton-proton
cycle (Bahcall 1989), and helium burning via the triple-alpha reaction
(Kippenhahn \& Weigert 1990).
During the later stages of the pre main sequence
evolution in the cases ``with capture'', the DS becomes dense enough
to capture DM from the ambient medium via elastic scattering.  Already
before fusion can begin, and possibly again after the onset of fusion,
captured DM can provide an important energy source with accompanying
luminosity 
\be\label{caplum}
L_{cap}=2m_{\chi}\Gamma_{cap}=2m_{\chi}f_Q\int dV
\rho_{cap}^2\sv/m_{\chi} \ee
and again $f_Q = 2/3$.

\section{Results of Stellar Structure Analysis}

Using our polytropic model for dark stars, we have started with
3$\msun$ stars and allowed baryonic matter to accrete onto them until
they become
supermassive with $M_* >10^5 \msun$.
We display results for the case without capture (but with extended 
adiabatic contraction) as well as the case with
capture for a variety of WIMP masses $m_\chi$ = 10 GeV, 100 GeV, and 1
TeV.  We have run models for a variety of accretion rates of baryons
onto the star including constant accretion rates of $\dot M = 10^{-1},
10^{-2}, 10^{-3} \msun$/yr.
We will present results for $\dot M = 10^{-3} \msun$/yr, which is
approximately the average rate calculated by Tan \& McKee (2004) and
by O'Shea \& Norman (2007).

\begin{figure}[t]
\includegraphics[width=0.98\textwidth]{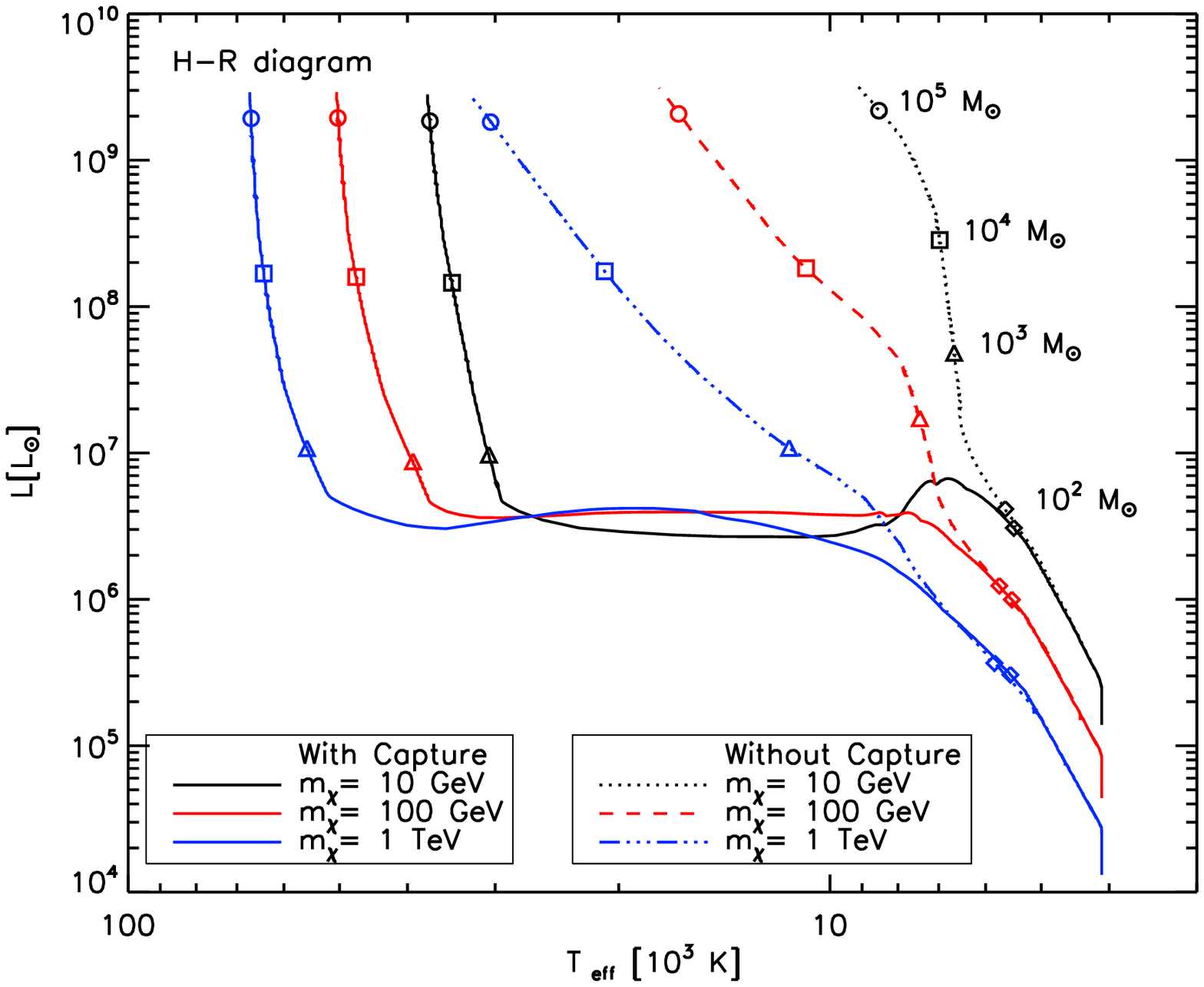}
\\
\caption{Hertzsprung-Russell diagram for dark stars for accretion rate
  $\dot M = 10^{-3} \msun$/yr.  and a variety of WIMP masses as
  labeled for the two cases: (i) ``without capture'' but with extended
  adiabatic contraction (dotted lines) and (ii) ``with capture''
  (solid lines).  The case with capture is for product of scattering
  cross section times ambient WIMP density $\sigma_c \bar\rho_\chi =
  10^{-39} {\rm cm}^2 \times 10^{13}$GeV/cm$^3$.  Also labeled are
  stellar masses reached by the DS on its way to becoming
  supermassive. The final DS mass was taken to be $1.5\times 10^5
  \msun$ (the baryonic mass inside the initial halo), but could vary
  from halo to halo, depending on the specifics of the halo mergers.
}
\label{dens}
\end{figure}

\begin{figure}[t]
\includegraphics[width=0.98\textwidth]{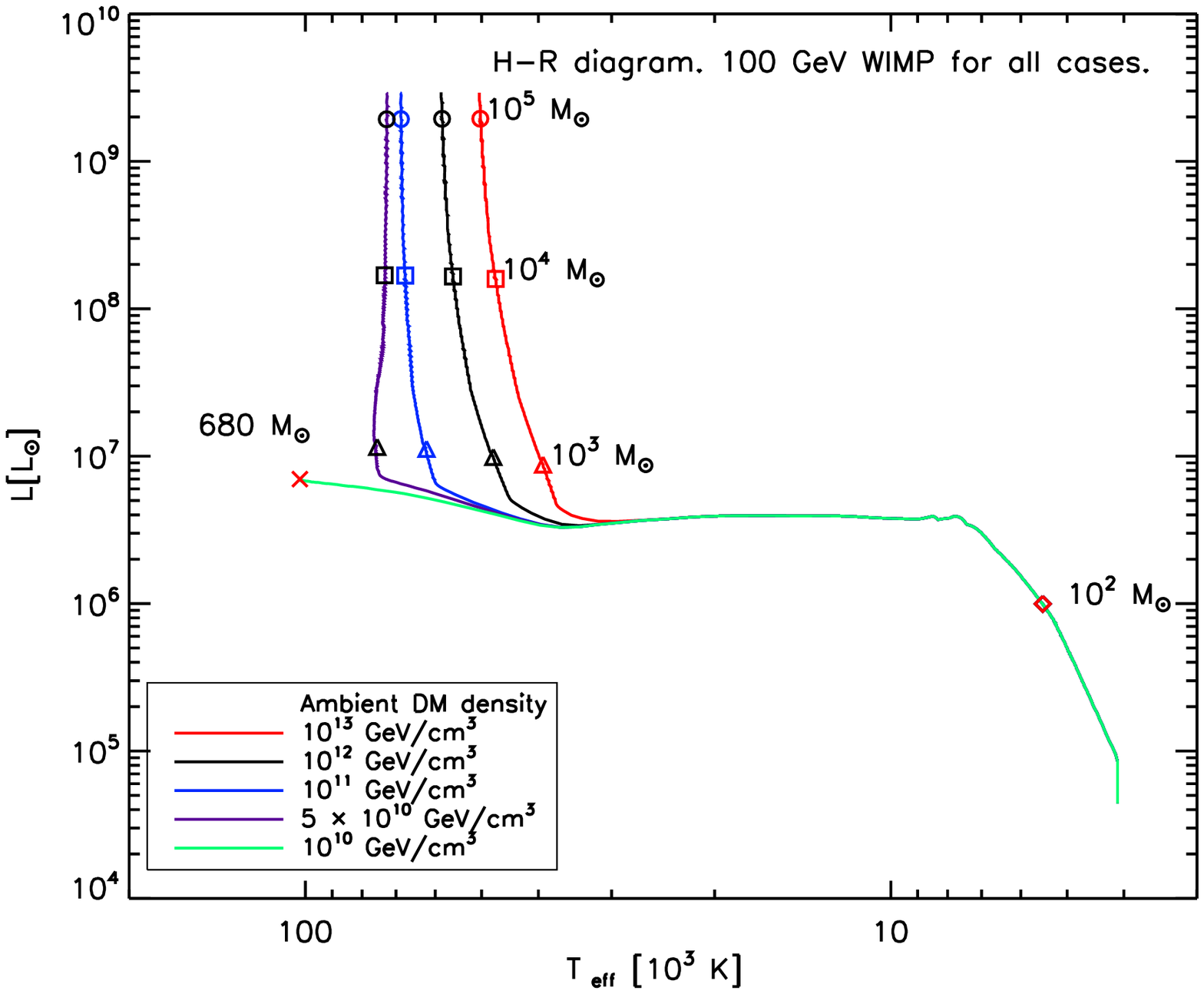}
\\
\caption{Hertzsprung-Russell diagram for DS for the case ``with
  capture'' for 100 GeV WIMP mass and accretion rate $\dot M = 10^{-3}
  \msun$/yr.  The different curves are for a variety of ambient DM
  densities $\bar \rho_\chi$ as labeled for scattering cross section
  $\sigma_{c} = 10^{-39} {\rm cm}^2$.  The results depend only on the
  product $\sigma_c \bar\rho_\chi$ so the different curves could
  equivalently refer to different $\sigma_{c}$ for a given
  $\bar\rho_\chi$.  }
\label{DM}
\end{figure}

Our stellar evolution results can be seen in the Hertzsprung-Russell diagram of
Figure 1 for the case of a $10^6 \msun$ halo.  The dark star travels up to increasingly
higher luminosities as it becomes more massive due to accretion. We
have labeled a sequence of ever larger masses until all the baryons
(150,000$\msun$) in the original halo are consumed by the SMDS.  As
the mass increases, so do the luminosity and the surface
temperature. In the cases ``without capture,'' the radius increases
continuously until all the baryons have been eaten.  In the cases
``with capture'', we have taken the (overly conservative) assumption
that the DM from adiabatic contraction is depleted after $\sim
300,000$ yrs as in our earlier papers; then the luminosity plateaus
for a while while the DS contracts until eventually it is dense enough
to capture further DM.
  
We note that, for the case ``without capture'', the H-R diagram is
unchanged by varying the accretion rate: only the time it takes to get
from one mass stage to the next changes, but the curves we have
plotted apply equally to all accretion rates. Similarly, given
$m_\chi$, the following quantities are the same regardless of
accretion rate: $R_*, T_{eff}, \rho_c,$ and $T_c$.
  
In a beautiful  paper, Hoyle \& Fowler (1963) studied supermassive
stars in excess of $10^3 \msun$ and found results germane to our
work. They treated these as $n=3$ polytropes (just as we do) dominated
by radiation pressure, and found the following results: $R_* \sim
10^{11} ({M_* / \msun})^{1/2} (T_c/10^8{\rm K})^{-1} {\rm cm}$, ${L_*
  / L_\odot} \sim 10^4 {M_* / M_\odot}$, and $T_{eff} \sim 10^5
(T_c/10^8{\rm K})^{1/2}{\rm K}$.  While some of the details of their
calculations differ from ours, taking the central temperature
appropriate to DS in the above relations roughly reproduces our
results (to O(1)). For example, by using the temperature appropriate
to dark stars with extended AC ($T_c \sim 10^6$K) rather than the much
higher central temperature ($T_c > 10^8$K) appropriate to nuclear
power generation, the above relations  show that DS have much
larger radii and smaller surface temperatures than fusion powered
stars.  We wish to draw particular attention to the fact that
luminosity scales linearly with stellar mass, and is independent of
power source.
  
Figure 2 plots the H-R diagram ``with capture'' for a single WIMP mass
of 100 GeV, for $\dot M = 10^{-3} \msun$/yr, and for $\sigma_c =
10^{-39}$cm$^2$, but for a variety of ambient densities ranging from
$\bar \rho_\chi = (10^{10} - 10^{14})$GeV cm$^{-3}$.  The latter is
the density one finds due to adiabatic contraction at the photosphere
of the DS, and seems the largest sensible starting point for the value
of the ambient density.  Our previous paper (Spolyar et al. 2009)
considered the case of minimal capture with $10^{10}$GeV/cm$^3$, which
was artificially chosen so that the growth of the DS ceases at
680$\msun$, the radius shrinks, and then fusion and DM heating play
equal roles.  For ambient densities below $5 \times
10^{10}$GeV/cm$^3$, the DS mass growth shuts off well before the star
becomes supermassive for the following reason.  The cases with capture
all take place at higher stellar densities than the cases without; the
density must be high enough to be able to capture WIMPs.  Consequently
the surface temperature is larger and accretion is shut off more
easily by radiation coming from the star.  The case of ambient
density $10^{10}$GeV/cm$^3$ (from our previous paper) is a very
carefully chosen (delicate) situation. On the low side of this
density, DM heating is completely irrelevant and fusion tells the
whole story; on the other hand, for any density $\bar \rho_\chi \geq
{\rm few} \times 10^{10}$ GeV/cm$^3$, DM heating is so dominant over
fusion that the DS can just continue growing in mass. At these higher
densities the surface temperature never becomes hot enough ($\approx
100,000$ K) for feedback effects from the star to cut off
accretion. Between 50,000 K and 100,000 K feedback effects are
included, and they act to reduce the accretion rate, but they never
shut it off entirely for densities above $5 \times 10^{10}$GeV/cm$^3$.
In reality a star that is moving around can sometimes hit pockets of
high $\bar\rho_\chi$ (where it is DM powered and grows in mass) and
sometimes hit pockets of low $\bar\rho_\chi$ (where fusion takes
over).  As long as the ambient density remains at least this large,
the star can reach arbitrarily large masses and eat the entire
baryonic content of the halo.
  
As described previously, the capture mechanism depends on
 the product of scattering cross section times ambient
density, $\sigma_c \bar \rho_\chi$, rather than on either of these
quantities separately.  Hence our current discussion
could trade off ambient density vs. cross section. For example, the
product is the same for $\bar \rho_\chi = 5 \times 10^{10}$ GeV/cm$^3$
and $\sigma_c = 10^{-39}$ cm$^2$ as it is for $\bar \rho_{\chi} =
10^{14}$GeV/cm$^3$ and $\sigma_c = 5 \times 10^{-43}$ cm$^2$.  Thus,
for the highest reasonable ambient density, the scattering cross
section can be several orders of magnitude lower than the experimental upper bound for
spin-dependent (SD) scattering and still provide substantial capture in DS. 
While the required $\sigma_c$ is ruled
out (Ahmed et al. 2009) for SI scattering at $m_\chi = 100$GeV, it is
below the bounds at low masses $m_\chi \sim 10$ GeV and in this 
case can lead to significant DM capture in the stars.
  
Above $\sim 100 \msun$, one can see that the stellar luminosity scales
as $L_* \propto M_*$ and is the same for all models for a given
stellar mass; this statement is essentially true for all stars no
matter the power source.  The reason is that at these masses, the star
is essentially radiation pressure supported throughout\footnote{There
is a slight deviation for the $10^3\msun$ case without capture,
where the star is still only 78\% radiation pressure supported with
the remaining pressure due to gas.}.  This same scaling in supermassive stars
was already noticed
by Hoyle and Fowler (1963).  The luminosity essentially
tracks (just below) the Eddington luminosity which scales as $L
\propto M_*$.
  
The curves with higher values of WIMP mass $m_\chi$ lie to the left of
the curves with lower $m_\chi$.  This can be understood as follows.
The DM heating rate in Eqn.(1) scales as $Q \propto \rho_\chi^2 /
m_\chi$.  Hence to reach the same amount of heating and achieve the
same luminosity, at higher $m_\chi$ the DS must be at higher WIMP
density, i.e., the stellar radius must be smaller, the DS is hotter,
and the corresponding surface temperature $T_{eff}$ is higher.  Also,
for higher $m_\chi$ the amount of DM in the star is smaller since the
star is more compact for the same number of baryons but $\rho_\chi
\propto n^{0.8}$ where $n$ is the hydrogen density.

Tables 1 and 2 show various stellar properties for a DS that forms in
a $10^6 \msun$ DM halo, as the star evolves to higher mass for the
case of $m_\chi=100$ GeV, for the two cases ``without'' and ``with''
capture respectively.
While the DM density is a gently decreasing function of radius for the
case without capture, it is extremely sharply peaked at the center of
the DS for the case ``with capture''.

One can see that, in the case ``without capture'', $\sim 10^4\msun$ of
dark matter must be annihilated away in order for the DS to reach
$10^5\msun$ for accretion rate $\dot M = 10^{-3} \msun$.  The time to
reach this mass is $\sim$ 100Myr.  For an alternative faster accretion
rate $\dot M = 10^{-2}\msun$, a smaller amount of DM must be
annihilated away, $\sim 100\msun$; then it takes the DS $\sim$ 10 Myr
to reach $\sim 10^5\msun$.  The caveat is that the DM orbits must
continue to penetrate into the middle of the DS for this length of
time in order for the DM abundance and consequent heating to continue;
it is the DM heat source that keeps the DS cool enough to allow it to
continue to accrete baryons.  Additionally, the assumption that
baryons continue to accrete onto the DS must continue to hold.  Yet,
in the time frame required, the original $10^6\msun$ halo will merge
with its neighbors, so that both the baryon and DM densities are
disturbed. These mergers could effect the DS in one of two ways:
either they provide more baryons and DM to feed the SMDS so that it
ends up being even larger, or they disrupt the pleasant high DM
environment of the SMDS so that it loses its fuel and converts to an
entirely fusion powered star.  Continued growth of the DS is quite
plausible since simulations with massive BHs in mergers show that they
prefer to sit close to the center of the density distribution or find
the center in a short time after the merger.

Someday detailed cosmological simulations will be required to answer
this question. Individual DS in different haloes may end up with a
variety of different masses depending on the details of the evolution
of the haloes they live in.  The case studied in this paper is clearly
a simplistic version of the more complicated reality, but illustrates
the basic idea that supermassive stars may be created by accretion
onto DS, either with or without capture.

 \begin{table*}
 \caption{ Properties and evolution of dark stars for $m_\chi = 100$
   GeV, $\dot M = 10^{-3} M_\odot/{\rm yr}$ for the case without
   capture but with extended adiabatic contraction. The DM halo was
   taken to be at a redshift of 20 with a concentration parameter of
   3.5 and with a mass of $10^6{\rm M_\odot}$. Shown are the stellar
   mass $M_*$, the DS luminosity $L_*$, the stellar radius $R_*$, the
   surface temperature $T_{eff}$, the central baryon density $\rho_c$,
   the central temperature $T_c$, the amount of DM in the DS $M_\chi$
   (due to both adiabatic contraction and capture), the central WIMP
   density $\rho_{\chi,c}$, and the amount of DM consumed by the DS
   $M_{Ann}$.
}
\begin{center}
{
\small
 \begin{tabular}{||c|c|c|c|c|c|c|c|c||}
 \hline\hline

$M_*$ & $L_*$ & $R_* $ & $T_{eff} $ & 
$\rho_{c} $ & $T_c $ & $M_{\chi}$ 
& $\rho_{\chi,c}$ & $M_{\rm Ann}$\\ 
$(\rm M_\odot)$ & $(10^6 L_\odot)$ & $({\rm AU})$ & $ (10^3 {\rm K})$ & 
$({\rm g/cm}^3)$ & $(10^5 {\rm K})$ & ($\rm M_\odot$) 
& (g/cm$^3$) & ($\rm M_\odot$) \\ 
\hline
10.1
& $0.13$
&$3.1$
&4.3
&$2.8\times10^{-7}$
&1.08
&0.02
&$9.2\times10^{-10}$
&$7\times10^{-5}$\\
\hline
  100
& $1.2$
& $5.2$
& $5.7$
& $7.4\times10^{-7}$
& $3.4$
& $0.1$
& $1.5\times 10^{-9}$ 
& $5.6\times10^{-3}$
\\
\hline
  500
& $9.7$
& $9.3$
& $7.2$
& $4.3\times10^{-6}$
& $8.3$
& $0.5$
& $5.8\times 10^{-9}$
&$0.26$
\\
\hline
  $10^3$
& $17$
& $12$
& $7.5$
& $4.6\times10^{-6}$
& $9.8$
& $0.84$
& $3.3\times 10^{-10}$ 
& $0.9$
\\
\hline
  $10^4$
& $182$
& $18$
& $10.8$
& $1.3\times10^{-5}$
& $21$
& $5.3$
& $8.4\times 10^{-9}$ 
& $86$
\\
\hline
  $10^5$
& $2100$
& $26$
& $16.5$
& $4.1\times10^{-5}$
& $46$
& $31.2$
& $1.6\times 10^{-8}$
& $10750$
\\
\hline 
 \hline\hline
 \end{tabular} 
 \label{tab:ExpParam}
}
\end{center}
\end{table*}

 \begin{table*}
  \caption{ Properties and evolution of dark stars for case ``with
    capture'', for $m_\chi = 100$ GeV, $\dot M = 10^{-3} M_\odot/{\rm
      yr}$, and product of scattering cross section times ambient DM
    density $\sigma_c \bar{\rho}_{\chi}= 10^{-39} {\rm cm}^2 \times
    10^{13}$ GeV/cm$^3$.  The Halo has the same parameters as in Table
    1.  The quantities tabulated are the same as in Table 1.  The
    double horizontal line delineates the transition from
    adiabatically contracted DM to captured DM once the DS reaches
    $\sim 1000\msun$ (after this point, the DM from AC has been
    annihilated away).
}

\begin{center}
{
\small
 \begin{tabular}{||c|c|c|c|c|c|c|c|c||}
 \hline\hline

$M_*$ & $L_*$ & $R_* $ & $T_{eff} $ & 
$\rho_{c} $ & $T_c $ & $M_{\chi}$ 
& $\rho_{\chi,c}$ & $M_{\rm Ann}$ \\ 
$(M_\odot)$ & $(10^6 L_\odot)$ & $({\rm AU})$ & $ (10^3 {\rm K})$ & 
$({\rm g/cm}^3)$ & $(10^5 {\rm K})$ & ($\rm M_\odot$) 
& (g/cm$^3$)  & ($\rm M_\odot$)  \\ 
\hline
10.1
&0.13
&3.1
&4.3
&$2.8\times10^{-7}$
&1.08
&0.02
&$9.2\times10^{-10}$
&$4.0\times10^{-5}$
\\
\hline
  100
& $1.2$
& $5.1$
& $5.7$
& $7.4\times 10^{-7}$
& $3.5$
& $0.1$
& $1.3\times 10^{-9}$
& $2.7\times10^{-3}$ \\
\hline
  500
& $5.5$
& $6.0$
& $7.8$
& $1.6\times 10^{-5}$
& $13$
& $0.3$
& $1.6\times 10^{-9}$
& $0.09$ \\
\hline\hline\hline
  $10^3$
& $8.8$
& $0.3$
& $39$
& $2.9\times 10^{-1}$
& $390$
& $3.1\times 10^{-6}$
& $5.4\times10^{-7}$
& $0.27$ \\
\hline
  $10^4$ 
& $161$
& $0.9$
& $47$
& $1.1\times 10^{-1}$
& $440$
& $2.9 \times 10^{-5}$ 
& $1.1\times10^{-6}$
& $77 $ \\
\hline
  $10^5$ 
& $1950$
& $2.7$
& $50$
& $3.8\times 10^{-2}$
& $450$
& $1.3\times 10^{-4}$
& $3.0\times10^{-6}$
& $9900$
\\
 \hline 
 \hline\hline
 \end{tabular} 
 \label{tableCap}
}
\end{center}
\end{table*}

Tables 1 \& 2 present models for DSs which form in DM halos of mass
for a $10^6 \msun$, but SMDS could also form in a variety of halo
masses with different final stellar masses.  For example, a
hydrogen/helium molecular cloud may start to contract in a $10^8
\msun$ halo and produce a DS.  Here the situation is more complicated.
The virial temperature of the halo exceeds $10^4$K, the surface
temperature we have found for a DS in equilibrium.  Hence it is not
clear how accretion onto the DS will proceed. This is the subject of
future work.  The accretion is expected to be faster due to the
increased ambient temperature.  We extended our models to $10^8\msun$
halos in which a more extended period of accretion can lead to SMDS
with even larger masses.  Tables 3 \& 4 show examples of ``without''
and ``with'' capture cases respectively for $10^8\msun$ halos. The
baryonic mass in the halo is $1.5 \times 10^7 \msun$. Potentially all
of this mass could go into the SMDS.  Then once the DM runs out it
becomes an enormous Pop III fusion powered star, which soon burns out
and becomes a black hole with mass $>10^7 \msun$.  Such a large BH at
early times would clearly help to explain the many large BH found in
the Universe at early times and today.  In addition, since $L_*
\propto M_*$, we can predict that the luminosity of the $M_* \sim
10^7\msun$ SMDSs would be $L_* \sim 10^{11}L_\odot$, even easier to
detect than a $10^5$ SMDS.

\begin{table*}
\caption{ Properties and Evolution of dark stars for $m_\chi = 100$
  GeV, $\dot M = 10^{-1} M_\odot/{\rm yr}$ for the case without
  capture but with extended adiabatic contraction. The DM halo was
  taken to be at a redshift of 15 with a concentration parameter of
  3.5 and with a mass of $10^8{\rm M_\odot}$.  The
  quantities tabulated are the same as in Table 1.
}

\begin{center}
{
\small
 \begin{tabular}{||c|c|c|c|c|c|c|c|c||}
 \hline\hline

$M_*$ & $L_*$ & $R_* $ & $T_{eff} $ & 
$\rho_{c} $ & $T_c $ & $M_{\chi}$ 
& $\rho_{\chi,c}$ & $M_{\rm Ann}$ \\ 
$(M_\odot)$ & $(10^6 L_\odot)$ & $({\rm AU})$ & $ (10^3 {\rm K})$ & 
$({\rm g/cm}^3)$ & $(10^5 {\rm K})$ & ($\rm M_\odot$) 
& (g/cm$^3$)  & ($\rm M_\odot$)  \\ 
\hline
12
&0.19
&3.6
&4.3
&$1.6\times10^{-7}$
&0.90
&0.03
&$8.4\times10^{-10}$
&$1.1\times10^{-6}$
\\
\hline
  100
& $1.9$
& $6.5$
& $5.7$
& $3.8\times 10^{-7}$
& $2.7$
& $0.2$
& $1.3\times 10^{-9}$
& $7.6\times10^{-5}$ \\
\hline
  $10^3$
& $23$
& $15$
& $7.1$
& $2.3\times 10^{-6}$
& $7.8$
& $1.4$
& $4.0\times10^{-9}$
& $1.2\times 10^{-2}$ \\
\hline
  $10^4$ 
& $172$
& $28$
& $8.6$
& $3.5\times 10^{-6}$
& $14$
& $9.7$
& $4.3\times10^{-9}$
& $0.9 $ 
\\
\hline
 $10^5$
& $2100$
& $39$
& $14$
& $1.3\times 10^{-5}$
& $31$
& $56$
& $9.1\times10^{-9}$
& $109$
\\
\hline
$10^6$
&$2.2\times 10^4$
&61
&19
&$3.3\times10^{-5}$
&64
&355
&$1.5\times10^{-8}$
&$1.1\times10^4$
\\
\hline
$10^7$
&$2.2\times10^5$
&97
&27
&$8.3\times10^{-5}$
&127
&2200
&$2.3\times10^{-8}$
&$1.2\times10^6$
\\
 \hline 
 \hline\hline
 \end{tabular} 
 \label{tableCap}
}
\end{center}
\end{table*}

\begin{table*}
  \caption{ Properties and evolution of dark stars for case ``with
    capture'', for $m_\chi = 100$ GeV, $\dot M = 10^{-1} M_\odot/{\rm
      yr}$, and product of scattering cross section times ambient DM
    density $\sigma_c \bar{\rho}_{\chi}= 10^{-39} {\rm cm}^2 \times
    10^{13}$ GeV/cm$^3$.  The DM halo has the same parameters as in
    Table 3.  The quantities tabulated are the same as in Table 1.
    The double horizontal line delineates the transition from
    adiabatically contracted DM to captured DM once the DS reaches
    $\sim 4\times 10^4\msun$ (after this point, the DM from AC has
    been annihilated away).
}

\begin{center}
{
\small
 \begin{tabular}{||c|c|c|c|c|c|c|c|c||}
 \hline\hline

$M_*$ & $L_*$ & $R_* $ & $T_{eff} $ & 
$\rho_{c} $ & $T_c $ & $M_{\chi}$ 
& $\rho_{\chi,c}$ & $M_{\rm Ann}$ \\ 
$(M_\odot)$ & $(10^6 L_\odot)$ & $({\rm AU})$ & $ (10^3 {\rm K})$ & 
$({\rm g/cm}^3)$ & $(10^5 {\rm K})$ & ($\rm M_\odot$) 
& (g/cm$^3$)  & ($\rm M_\odot$)  \\ 
\hline
11
&0.18
&3.64
&4.3
&$1.6\times10^{-7}$
&0.9
&0.03
&$8.4\times10^{-10}$
&$5.6\times10^{-7}$\\
\hline
100
&1.8
&6.5
&5.7
&$3.8\times10^{-7}$
&2.7
&0.2
&$1.3\times10^{-9}$
&$3.8\times10^{-5}$\\
\hline
$10^3$
&22
&14
&7.2
&$2.3\times10^{-6}$
&7.8
&1.4
&$3.6\times10^{-9}$
&$6.1\times20^{-3}$\\
\hline
$10^4$
&173
&23
&9.4
&$5.8\times10^{-6}$
&16
&8.3
&$2.9\times10^{-9}$
&0.44\\
\hline
$4.1\times10^4$
&740
&1.8
&49
&$5.7\times10^{-2}$
&444
&0.18
&$7.2\times10^{-9}$
&6.0\\
\hline\hline\hline
$10^5$
&$1.9\times10^{3}$
&2.7
&51
&$3.8\times10^{-2}$
&452
&$1.3\times10^{-4}$
&$2.9\times10^{-6}$
&91
\\
\hline
$10^6$
&$2.1\times10^4$
&8.5
&51
&$1.2\times10^{-2}$
&456
&$2.7\times10^{-5}$
&$1.5\times10^{-4}$
&$1.1\times10^4$\\
\hline
$10^7$
&$2.1\times10^5$
&27
&51
&$3.9\times10^{-3}$
&457
&$4.0\times10^{-10}$
&$1.0\times10^2$
&$1.1\times10^6$
\\
 \hline 
 \hline\hline
 \end{tabular} 
 \label{tableCap}
}
\end{center}
\end{table*}

{\it General Relativistic Instability:} The pulsational stability of
supermassive stars is an interesting issue.  They are
radiation-pressure dominated with adiabatic index close to $\gamma = 1
+ 1/n = 4/3$, the value that yields neutral stability to radial
pulsations for Newtonian bodies with no rotation.  Indeed general
relativistic corrections (which scale as $GM_*/R_*$) act in the
direction of destabilizing stars and are particularly important for
high mass stars.  Fowler (1966) examined the
stability of supermassive stars using polytropes with $n=3$ (see
Wagoner 1969 for a review).
Fowler found that, for the case of no rotation, radial oscillations
become dynamically unstable and prevent standard stars more massive
than $10^5\msun$ from reaching a phase of hydrogen burning before
collapse.
Yet he also found that a small amount of rotation can stabilize the stars,
so that rotating stars as heavy as $10^8 \msun$ could be stable en
route to reaching hydrogen burning.

In the case of DS, stability to radial pulsations is much easier to
achieve.  DS have much larger radii and lower temperatures than fusion
powered stars, so that the GR corrections $\sim GM_*/R_*$ are much
smaller.  The upper limit on the allowed stellar mass will be
larger. In any case, SMDSs are undoubtedly rotating, so that very
large stable masses can be achieved (even in the case of rotating
ordinary stars, the mass limit is $10^8 \msun$).  In the future we
suggest a stability analysis of our models.

\section{Detectability with JWST}

We discuss the capabilities of JWST to discover dark stars, following
the properties of the telescope described by Gardner et al. (2006,
2009).  The telescope is designed to be diffraction limited at a
wavelength $\lambda_{obs}=2\mu$m.  The Near Infrared Camera (NIRCam)
will operate in the wavelength range $\lambda =(0.6-5)\mu$m and the
Mid-Infrared Camera (MIRI) will operate in the wavelength range
$\lambda =5-27\mu$m.  In an exposure of duration $10^4$s, NIRCam will
have a limiting sensitivity of 11.4 nJy ($1\, {\rm nJy}= 1 \times 10^{-32}$
ergs~s$^{-1}~$cm$^{-2}~$H$z^{-1}$) in the $2\mu$m band, and 13.8nJy in
the $3.5\mu$m band and MIRI will have a sensitivity of $700\mu$Jy in
the $\lambda=10 \mu m$ band (in all cases limiting sensitivities are
for a S/N=10).  With longer exposure times the limiting flux
detectable will scale as $\sqrt{t}$ for the same S/N.  
DS will be characterized by black body spectra with surface
temperatures $T_{eff}$ $\lesssim$ $5\times 10^4$K.  In addition, DS
are also predicted to have  hydrogen lines.

We determine the detectability of dark stars located at various
redshifts z= 5, 10, and 15 using the standard Planck spectrum of
blackbody with surface temperature $T_{eff}$ and radius $R_*$ (for DSs
similar to those from Tables 1-4) in a cosmology with $H_0 = 74,
\Omega_\Lambda = 0.71, \Omega_M= 0.29$.

\begin{figure}[t]
\includegraphics[width=0.5\textwidth]{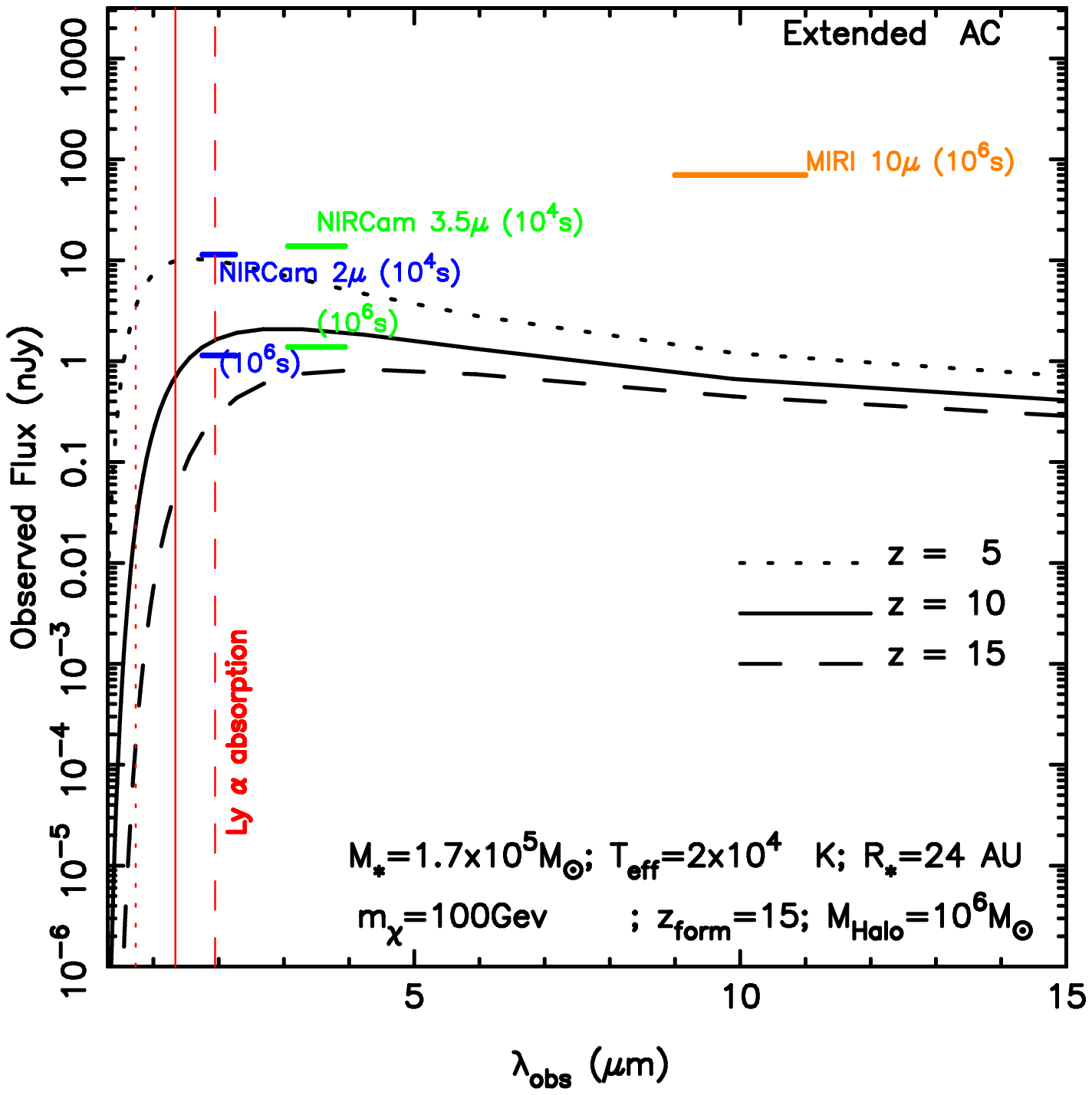}
\includegraphics[width=0.5\textwidth]{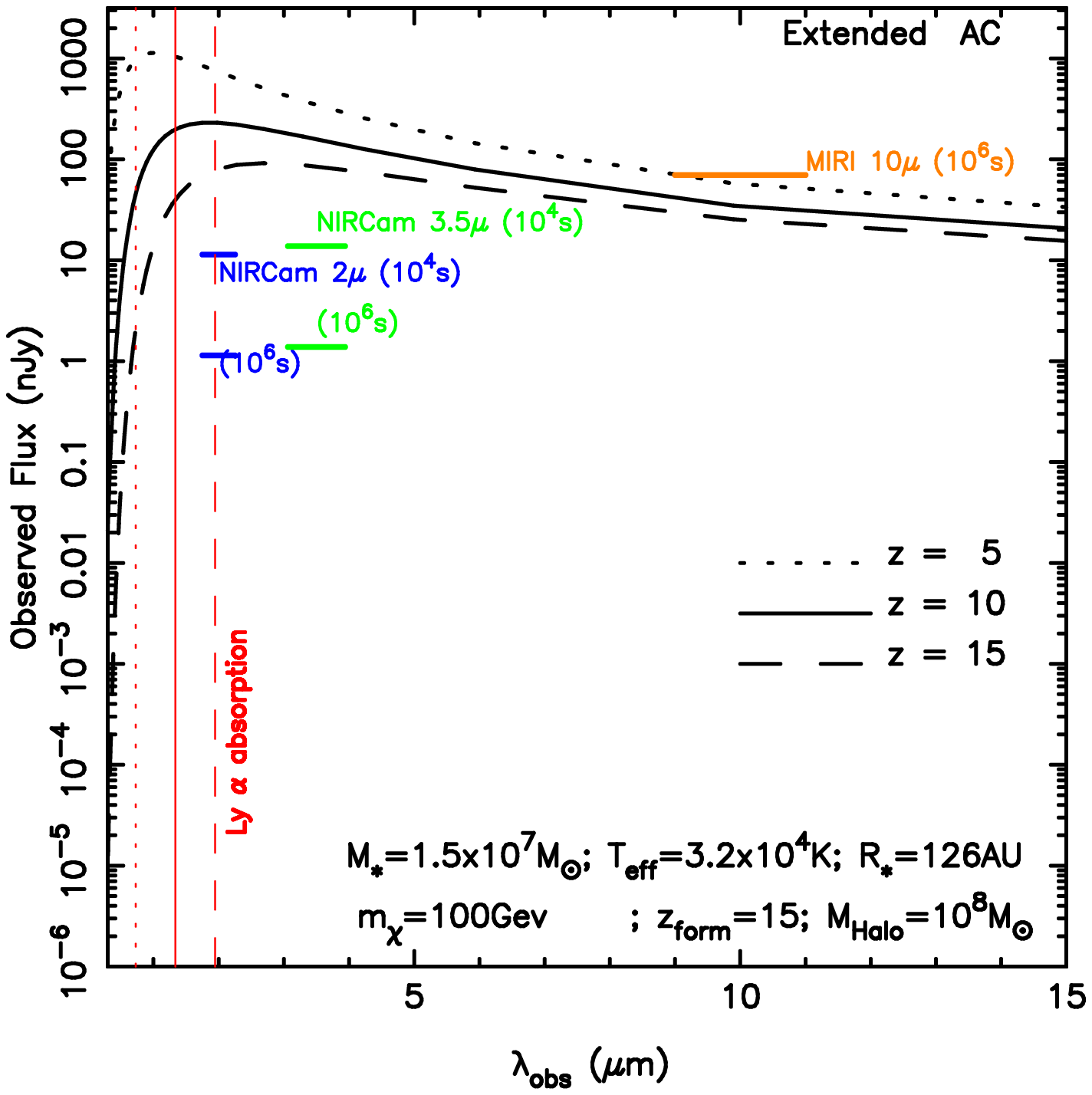}
 \caption{Black body spectra of two dark stars formed via extended
   adiabatic contraction (``without capture'') for $m_\chi$=100
   GeV. Left panel: $1.7\times10^5~\msun$ SMDS in a $10^6~\msun$ halo.
   Right panel: $1.5\times10^7~\msun$ SMDS in $10^8~\msun$ halo.  The
   black body flux is shown at $z=15$ (formation redshift) and at
   $z=10$ and 5 (see line legends) assuming that the dark star
   survives till the lower redshifts. Blue dashes show sensitivity
   limit and bandwidth of NIRCam $2\mu$ (R=4) while the green dashes
   show the sensitivity limit and band width of the NIRCam 3.5$\mu$
   (R=4) band. The upper and lower dashes show the sensitivity limits
   after exposure times of $10^4$s, $10^6$s respectively. The
   sensitivity of MIRI ($10\mu$, R=5) is shown for exposure time of
   $10^6$s (orange dash). All sensitivities are computed assuming a
   S/N=10.  The red vertical lines show the location of the 1216$\AA$
   line redshifted from the rest-frame wavelength of the star at each
   of the three redshifts. The observed flux to the left of the
   vertical lines will decrease relative to the black curves
   depending on the model assumed for IGM absorption up to the
   redshift of reionization.}
\label{jwst_flux_ac}
\end{figure}
Figure~\ref{jwst_flux_ac} shows the observed black body flux
distribution of two SMDSs formed at $z=15$ for a WIMP mass $m_\chi =
100$GeV for the case of extended AC (without capture). The star in the
left panel is formed in a $10^6\msun$ halo and the star in the right
panel is formed in a $10^8~\msun$ halo and their stellar (baryonic)
masses are $1.7\times10^5~\msun$ and $1.5\times 10^7~\msun$
respectively.  Curves are shown assuming the SMDS formed at $z=15$ and
survived to various redshifts, at which it is still producing
blackbody radiation.  The $1.7\times10^5~\msun$ star (left panel) will
be detectable by JWST (NIRCam) in an exposure of a million seconds,
but only if it survives intact till $z=10$.  The $1.5\times
10^7~\msun$ star (right panel) will be detectable even in a shorter
$10^4s$ exposure even at $z=15$ in both the 2$\mu$ and 3.5$\mu$ bands.
The star on the right may be marginally detectable in a million second
exposure in the 10$\mu$ band of MIRI. The relative flux levels in the
three different bands will be important for distinguishing these
objects from galaxies.
 
The curves are not corrected for Ly-$\alpha$ absorption by the IGM
 but the red vertical lines show the location of the 1216$\AA$ line
 redshifted from the rest-frame wavelength of the star at each of the
 three redshifts. Flux at wavelengths to the left of the redline at
 each redshift is expected to be absorbed to some extent by the
 IGM. Since the surface temperatures of our stars are $\simeq 10^4$K,
 the majority of the Ly$\alpha$ absorption ($\lambda_{rest} =1216
 \AA$) is expected to occur at wavelengths shorter than that at which
 the peak flux is emitted.  We note that for a DS at $z=15$ the
 Ly~$\alpha$ absorption line lies at 1.94$\mu$ - roughly in the middle
 of the NIRCam 2$\mu$ band. In this case the flux in this band will be
 reduced by about a factor of two but will still be well above the
 detection limit (in $10^6$s). A detailed calculation of the absorption
 by the intergalactic medium is outside the scope of the current
 paper.

 We studied numerous other cases without capture as well: a variety of
 WIMP masses (10Gev -2TeV) as well as various formation redshifts
 ($z_{\rm form} = 10-20$).  We found that DSs with masses up to $M_* \sim
 10^5 \msun$ forming in $10^6~\msun$ halos at $z=20$ and shining at
 that redshift will in general not be detectable for any value of
 $m_\chi$ (they will become detectable only if they survive to and
 shine at much lower redshifts). 
 
 The smaller DS with $M_* = 800 \msun$ discussed 
 in our previous work (Freese et al. 2008a; Spolyar et al. 2009) will not be 
 visible in JWST (see also
 Scott et al 2010); they are several orders of magnitude below the
 detection limit of JWST (in $10^6$s).

\begin{figure}[t]
\includegraphics[width=0.5\textwidth]{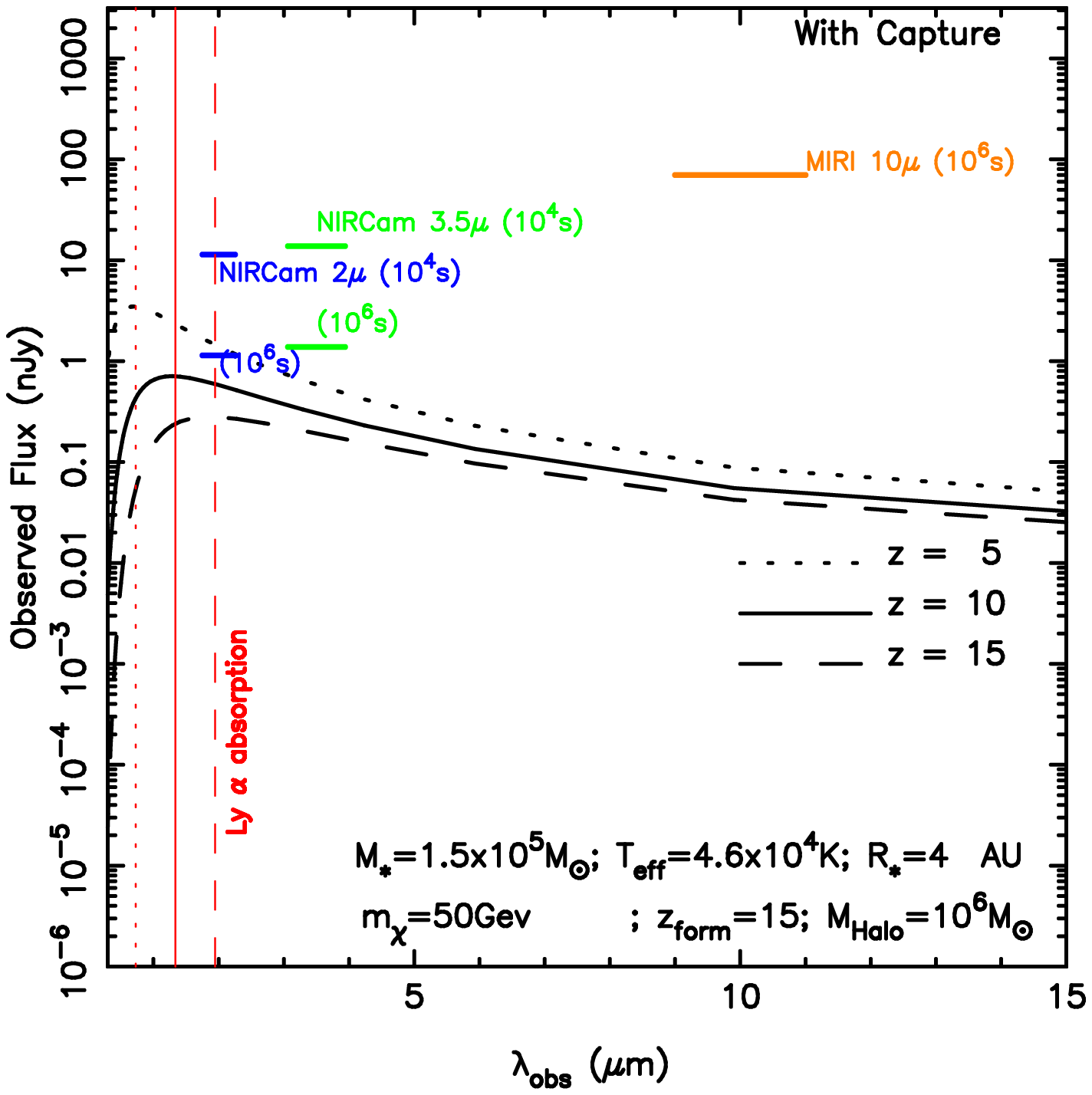}
\includegraphics[width=0.5\textwidth]{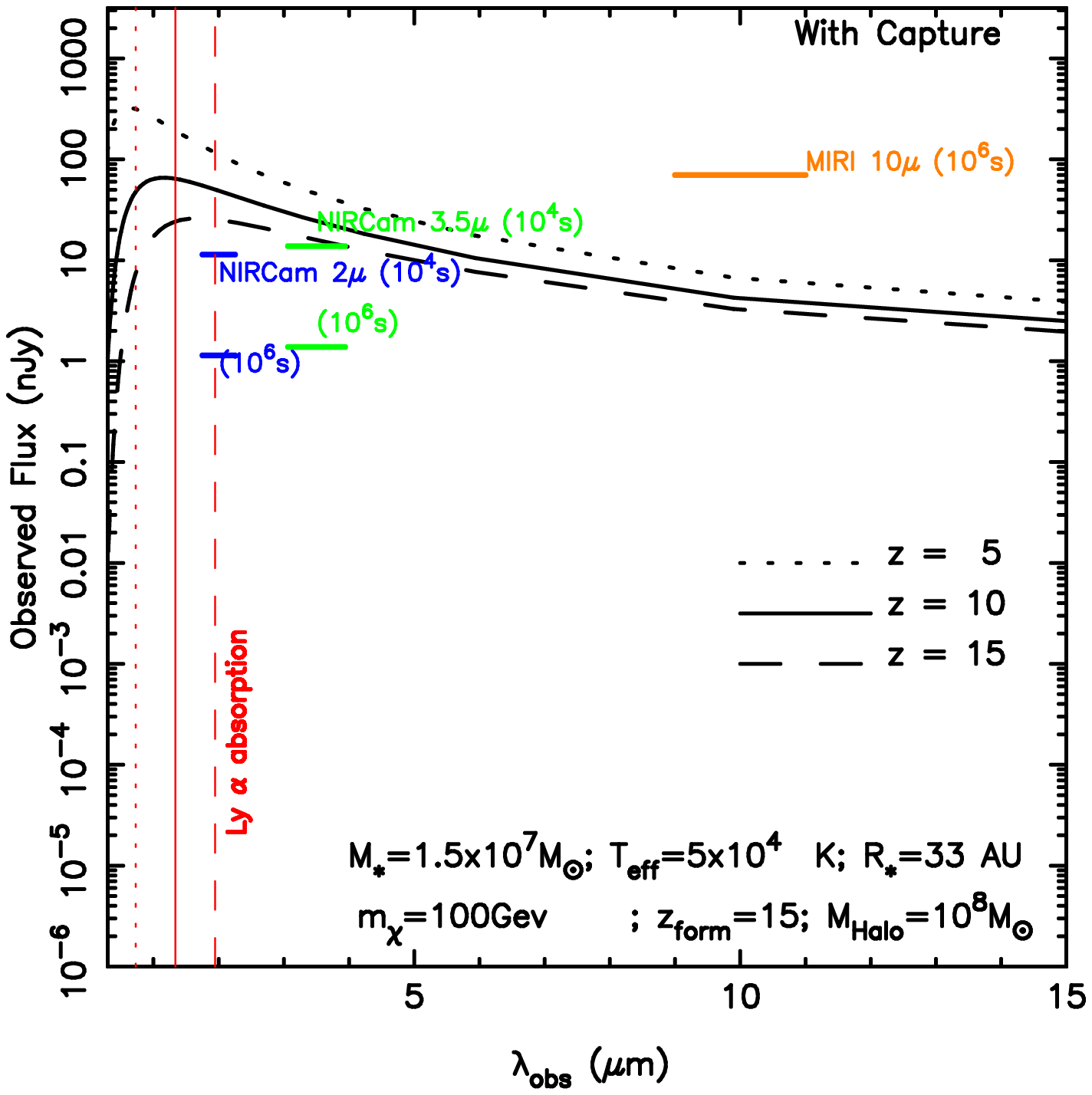}
\caption{Similar to Fig.~\ref{jwst_flux_ac} for dark stars formed ``with capture''. Left panel: $1.7\times10^5~\msun$ SMDS  formed in $10^6\msun$ halo ($m_\chi = 50$Gev).  Right panel:  $1.7\times10^7~\msun$ SMDS  formed in $10^8\msun$ halo ($m_\chi=100$Gev). }
\label{jwst_flux_cap}
\end{figure}

Figure~\ref{jwst_flux_cap} shows the observed black body flux
distribution of two dark stars formed at $z=15$ in halos of two
different masses for the case ``with capture''. Since DS formed via
capture are smaller (in radius) and hotter (than DS formed via
extended AC without capture), their peak wavelength tends to shift to
lower wavelengths, in some cases out of the range detectable by
JWST. $10^5 \msun $ stars in $10^6\msun$ halos are only detectable if
they survive until $z=5$ at which time they could be detectable in a
long ($10^6$s exposure).  On the other hand $10^7 \msun$ stars formed
in $10^8\msun$ halos will be easily detectable even in an exposure of
$10^4$s all the way out to $z=15$. For most other WIMP masses and
formation redshifts DSs formed in $10^6\msun$ halos via capture are
below the detection limit of JWST.

The prospect of detecting SMDS in JWST and confirming the existence of
a new phase of stellar evolution is exciting.  In the most optimistic cases,
detection in a number of different wavelength bands could be used to obtain a 
spectrum and differentiate
these dark stars from galaxies or other sources.

\section{Concluding Remarks}

Using our polytropic model for dark stars, we have 
considered accretion of baryonic matter onto the DS as they
become supermassive, $M_*>10^5 \msun$.  Such large
masses are possible because the dark star is cool enough (as long
as it is powered by DM) so that
radiative feedback effects from the star do not shut off the accretion
of baryons, as long as it is powered by DM. We considered two different scenarios for
supplying the required amount of dark matter: \hfil\break 1) {\bf The Case of
Extended Adiabatic Contraction, labeled ``without capture'' in the
figures.}  In this case dark matter is supplied by the gravitational
attraction of the baryons in the star. 
In triaxial haloes DM orbits are quite complex and the DM in the core
is harder to deplete than previously estimated.  This case does {\it
  not} include any captured DM, and relies solely on the particle
physics of WIMP annihilation.  To grow to a $10^5 \msun$ SMDS in
a $10^6 \msun$ halo, or to grow to a $10^7 \msun$ SMDS in a
$10^8 \msun$ halo, the amount of DM consumed can be as much as $\sim 1\%$
of the total DM in the halo (depending on the accretion rate).  
This amount is not unreasonable, since Valluri etal (2010)
found that the fraction of box and chaotic DM orbits is as high as 85\% 
in triaxial haloes and remains over 10\% when a  significant compact baryonic 
component causes the halo to become axisymmetric at small radii.
Future work will be required to
accurately obtain WIMP orbits, densities, and timescales (work in
progress).  For now we took the simplistic approach of
using our previous prescription for adiabatic contraction in a spherical potential but not
removing the annihilated DM.  
\hfil\break 2) {\bf The Case of Extended
Capture, labeled ``with capture'' in the figures.}  Here the original
DM inside the star from adiabatic contraction is assumed to be
depleted after $\sim$ 300,000 yrs, then the star begins to shrink
somewhat, and capture of DM from the surroundings takes place as it
scatters elastically off of nuclei in the star.  In this case the
additional particle physics ingredient of WIMP scattering is required.

In this paper we studied the formation of $10^5 \msun$ SMDS  in $10^6 \msun$
DM haloes and $10^7 \msun$ SMDS  in $10^8 \msun$ haloes.
These stars become very bright, $L_* \sim (10^9 - 10^{11}) L_\odot$.
Figure 1 shows the H-R diagram for a variety of WIMP masses, and
follows the dark star as it climbs up to ever higher masses.  
They live millions to billions of years, depending on the
merger history with other haloes.  Once the DM runs out, the SMDSs
have brief lives as fusion powered Pop III stars before collapsing
into $>10^5 \msun$ black holes, possible seeds for many of the big BH
seen in the Universe today and at early times.
 A proper study of the final mass of the DS and resultant BH will
depend on cosmological simulations.  The original halo containing the
DS will merge with other haloes.  No one knows what exactly will
happen to the DM density in the vicinity of the DS when this
happens. The DS could end up even more massive.  
DS may also form in larger haloes that form at later times, as long as
the baryonic content is still only H and He.  Localized regions with
this property could exist even at redshifts $z<7$ (Furlanetto \& Loeb
2005; Choudhury \& Ferrara 2007).  

SMDS would make plausible precursors of the $10^9 \msun$ black holes
observed at $z>6$ (Fan et al. 2003)
 of intermediate mass black holes;
of BH at the centers of galaxies; and of the BH inferred by
extragalactic radio excess seen by the ARCADE experiment (Seiffert et
al. 2009).  In addition, the BH remnants from DS could play a role in
high-redshift gamma ray bursts thought to take place due to accretion
onto early black holes

SMDSs could be detected by JWST for a variety of parameter ranges.
These are extremely bright objects $L_* \sim (10^9 - 10^{11}) L_\odot$
and yet are very cool $T \sim 10,000$K, so that their emitted light is in
the wavebands detectable by JWST.  The longer they live, the more easy they
are to detect.  Figures 3 and 4 give examples of what one could look for in JWST.
For the most optimistic cases, one could even test for the blackbody spectrum
in a number of different wavebands.  In principle hydrogen or helium lines
could be found to complement the blackbody emission. If, in addition, someday
high energy neutrinos are found to emanate from these stars, then it will
be  a clincher that DM annihilation took place inside the DS.

It is interesting to speculate that the Initial Mass Function of Population III fusion
powered  stars may be determined by the cutoff of the DM supply, which
may vary from one DS to another.  
Dark stars continue to accrete mass as long as the dark matter annihilation powers
the star and keeps it cool enough.  Once the DM fuel supply is exhausted, the star
shrinks and heats up, fusion begins, and the mass growth of the star is quickly halted
due to feedback from hot emitted photons.  Hence the details of the cutoff of the DM
supply may determine the sizes of Population III stars entering the fusion era.
The cutoff will take place at different DS masses in different haloes,
depending on the details of the cosmological merger history.
Different final DS masses may result for different individual DS
depending on the evolution of their parent haloes.

\section*{Acknowledgements}

We acknowledge support from: the DOE and MCTP via the Univ.\ of
Michigan (K.F. and C.I.);  DOE at Fermilab (D.S.); NSF grant AST-0908346 (M.V). K.F. is grateful to M. Begelman, W. Freedman,
J. Friedman, O. Gnedin, B. O'Shea, B. Schmidt,  P. Shapiro,
T. Tyson, and S. White for helpful discussions. We thank Jon Gardner for very
helpful comments regarding JWST.

\end{document}